# Molecular Structures and Strength-Toughness of Silica and Quartz

Xiaozhi Hu

Department of Mechanical Engineering+6

University of Western Australia

Perth, WA6009, Australia

xiao.zhi.hu@uwa.edu.au

The molecular structures of amorphous silica and crystalline quartz are used to predict their intrinsic strength and fracture toughness together with the theoretical strength of silicon dioxide $SiO_2$. At the atomic scale, the amorphous silica is characterized by the $SiO_2$ molecules or "crystals" around 0.31 nm in diameter. Thus, both silica and quartz can be modelled as crystalline materials. Theoretical predictions are confirmed by experiments data in literature.

Silicon dioxide ($SiO_2$), as one of the most abundant minerals found on earth, has been well studied [1-3] and molecular structures of amorphous silica glass and crystalline quartz have been known since 1932 [1]. While playing a significant role in various aspects of modern-day life, $SiO_2$ still generates new challenges in science and advanced technological industries, e.g., ultra-thin films for potential atomic sieves [4] and for chemobiotic catalysis on living cells [5], and high-pressure applications [6,7].

Nowadays, the molecular structures of amorphous silica glass and crystalline quartz in Fig. 1(a) and (b) are frequently used in research publications and textbooks [8]. "There is a beauty in the atomic continuous random networks that is irresistible and hard to deny - identical local units strung together to form a disordered whole … [3]". In this letter, the identical local (molecular) unit circled in Fig. 1(a) is treated as a "molecular" particle or crystal so that both amorphous silica and crystalline quartz can be modelled based on their characteristic molecular structure $C_{ch}$.

It may be, to the surprise of many, that the well-known molecular structures in Fig. 1(a) and (b) have never been linked explicitly and mathematically to the intrinsic tensile strength and fracture toughness of silica and quartz. To a certain degree, this can be understood because of the variation and unpredictability in the local atomic density fluctuation [9] and the ring size [10] in amorphous silica. This letter reports a simple model that can predict the intrinsic tensile strength and fracture toughness of amorphous silica and crystalline quartz based on their characteristic molecular structure $C_{ch}$, as circled in Fig. 1(a) and (b). The only extra information required is the theoretical tensile strength $\sigma_{th}$ around 16 or 17 GPa (linked to the atomic bonds) from the literature [11,12].

A simple and predictive theory is needed as direct measurements of fracture properties of silica and quartz commonly involve large scatters due to uncertainties in sample preparations (relevant to nano-/micro-scale surface defects) and adopted models. For instance, the fracture toughness $K_{IC}$ of quartz has recently been measured and estimated using 7 different indentation formulas under different indentation loads [13], showing $K_{IC}$ varies from 0.5 to 2.5 MPa√m. While the tensile strength of 1 nm silica fibres can be as high as 13.2 GPa, the strengths of silica nano-fibres with diameters > 40 nm can be as low as 1 GPa [11].

The common molecular feature of amorphous silica and crystalline quartz in Fig. 1(a) and (b) is they all contain regular "identical local units" [3]. If we zoom in into their molecular structures from macro- to atomic scale, the first-encountered "identical local unit" is the $SiO_2$ molecule in amorphous silica and the regular 6-ring crystal structure in crystalline quartz. Therefore, the characteristic molecular structure $C_{ch}$ = 0.31 nm for silica and $C_{ch}$ = 0.62 nm for quartz. In this way, the regular $SiO_2$ molecules in the amorphous silica glass can be considered as atomic-scale particles or "molecular crystals". The molecular structures of crystalline quartz and amorphous silica glass can be visualized by the simulated ice and water transition [14] shown in Fig. 1(c). Since these "molecular crystals" in amorphous silica can change their positions and orientations over a lengthy period because of glass flow and self-organizing of "molecular crystals", they possess some characteristics of "time crystals" [15,16].

The classic linear elastic fracture mechanics (LEFM), initiated by Griffith [17], for a shallow through-thickness crack $a$ in a large plate is as follows.

$$K_{IC} = 1.12 \cdot \sigma_f \sqrt{\pi \cdot a} \tag{1a}$$

$$\sigma_f = \frac{K_{IC}}{1.12\sqrt{\pi \cdot a}} \rightarrow \infty \quad \text{if } a \rightarrow 0 \tag{1b}$$

Although glass has been commonly believed to be an idea elastic material for brittle fracture specified by LEFM since Griffith's work in 1921 [17], Eq. (1b) shows there should be a lower limit for the surface crack $a$, below which LEFM is no longer valid. As illustrated in Fig. 1(e), LEFM overestimates the fracture strength $\sigma_f$ if the surface crack is around 3 times of the characteristic molecular structure $C_{ch}$ or shorter.

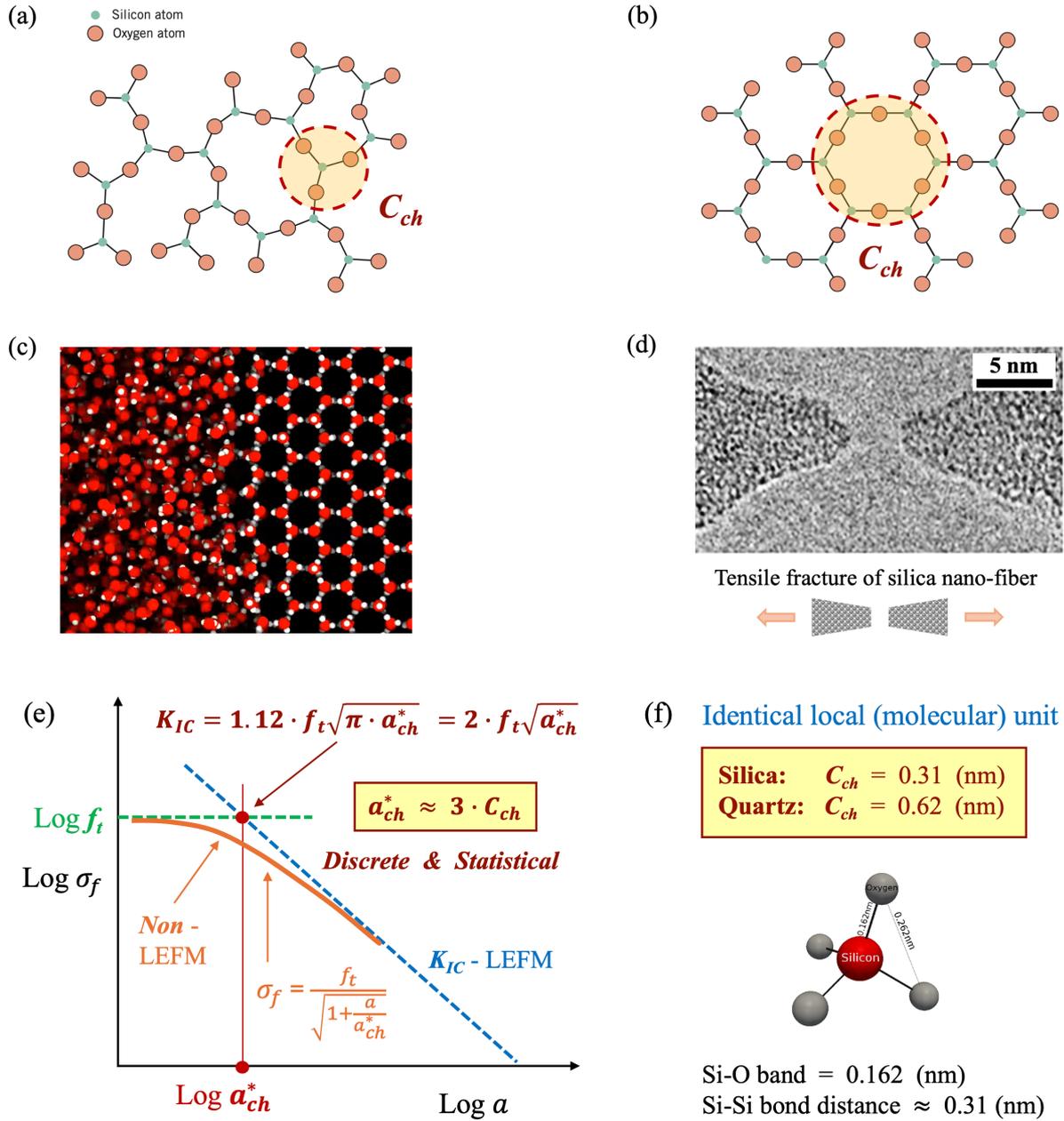

FIG. 1. (a) Two-dimensional (2D) amorphous silica glass with the circled identical local (molecular) unit $C_{ch}$ around 0.31 nm in diameter as in (f). (b) 2D crystalline quartz with circled "characteristic crystal structure" $C_{ch}$ around 0.62 nm in diameter. (c) Simulated water to ice transition and the change in molecular structures [14], akin to the molecular structures of silica and quartz. (d) TEM image of fractured silica nanofiber [11], indicating the characteristic molecular structures around 0.31 (nm) in amorphous silica. (e) LEFM is not valid for atomic or molecular scale cracks/defects. A non-linear asymptotic approximation (non-LEFM) is adopted for the relation between fracture stress $\sigma_f$ and crack $a$ with the scaling crack $a^*_{ch} = 3 \cdot C_{ch}$, which links the intrinsic strength $f_t$ and fracture toughness $K_{IC}$ criterion. (f) The identical local (molecular) unit or the characteristic molecular-structure $C_{ch} = 0.31$ nm for silica and $C_{ch} = 0.62$ nm for quartz, as circled in (a) and (b).

Fig. 1(e) shows an approximate asymptotic solution for atomic and molecular scale surface cracks, condensed from previous studies [18-20] when the condition pertinent to Eq. (1b) is relevant.

$$\sigma_f = \frac{f_t}{\sqrt{1+\frac{a}{a_{ch}^*}}} = \frac{f_t}{\sqrt{1+\frac{a}{3 \cdot C_{ch}}}} \quad (2)$$

$$K_{IC} = 2 \cdot f_t \sqrt{a_{ch}^*} = 2 \cdot f_t \sqrt{3 \cdot C_{ch}} \qquad \text{since } a_{ch}^* \approx 3 \cdot C_{ch} \quad (3)$$

The intersection of two distinct failure criteria, inert tensile strength $f_t$ and fracture toughness $K_{IC}$ of LEFM, defines a precise solution with $a = a_{ch}^*$ and $\sigma_f = f_t$.

The non-LEFM asymptotic approximation, Eq. (2), does not follow the strength or toughness criterion. Eq. (1a) of LEFM will overestimates the fracture strength $\sigma_f$ if the surface crack $a$ is not much larger than the characteristic crack $a_{ch}^*$, e.g. $0 \leftarrow a < 10 \cdot a_{ch}^*$. It is realized from the previous studied [e.g. 20], the characteristic crack $a_{ch}^*$ is linked to the characteristic microstructure $C_{ch}$, i.e., $a_{ch}^* = C \cdot C_{ch}$, where $C$ is a constant. A discrete and statistical analysis has then been used to model various experiments for $C$ = 0.5, 1.0, 1.5, 2.0, 2.5, 3.0, 3.5, 4.0 … [e.g. 21], showing $C$ = 3 provides the minimum standard deviation with a near perfect normal distribution. If $C$ < 3 or > 4, distributions are distorted away from the normal distribution together with larger standard deviations. It has been found [20] that $C$ = 3 provides good approximations for various solids with the characteristic microstructure $C_{ch}$ varying from the atomic scale (around 0.2 nm) all the way to macro-scale (around 10's mm).

The intersect of strength and toughness criteria $a_{ch}^*$ (= $3 \cdot C_{ch}$) in Fig. 1(e) has an important physical meaning, i.e., it is the width of the atomic-scale crack-tip damage/cracking zone with the damage zone length around $1.5 \cdot C_{ch}$ or between 1 and 2 $C_{ch}$ [20]. "Glass Breaks like Metal, but at the Nanometer Scale", reported in [22], is confirmed here by the non-LEFM model Eq. (2) via the characteristic molecular structure $C_{ch}$ (= 0.31 nm). As indicated by $a_{ch}^* = 3 \cdot C_{ch}$ in Fig. 1(e), the non-LEFM zone is around 1 to 2 nm.

It should be emphasized that both Eqs. (2) and (3) are predictive if the characteristic microstructure $C_{ch}$ and intrinsic strength $f_t$ are provided. Since the characteristic molecular structure $C_{ch}$ = 0.31 nm for silica and $C_{ch}$ = 0.62 nm for quartz, only the intrinsic strength $f_t$ needs to be determined for silica and quartz. The intrinsic strength of silica at room temperature has been estimated to be 10 – 12 GPa [23]. Eq. (3) predicts the fracture toughness $K_{IC}$ of silica with $C_{ch}$ = 0.31 nm is between 0.61 – 0.73 MPa√m, which is within the range of 0.59 – 0.78 MPa√m reported in literature [24].

The intrinsic strength $f_t$ of amorphous silica can be estimated using Eq. (2). A good quality silica may still contain distributed silicon-oxygen atom rings, e.g., with 4 to 9 silicon and 4 to 9 oxygen atoms [1, 3, 10]. Assuming the theoretical strength $\sigma_{th}$ = 16 – 17 GPa [11,12], the large "intrinsic molecular defect" $a_{th}$ (controlling fracture) may be in the range of 0.62 – 1.24 nm, i.e.,

$$f_t = \frac{\sigma_{th}}{\sqrt{1+\frac{a_{th}}{3 \cdot C_{ch}}}} = \frac{16-17}{\sqrt{1+\frac{0.62-1.24}{3 \cdot 0.31}}} = 10.5 - 13.2 \ (\text{GPa}) \tag{4}$$

Considering the uncertainty in the "intrinsic molecular defect" $a_{th}$, the intrinsic tensile strength $f_t$ of amorphous silica can be adjusted to a wider range of 10.0 – 13.5 GPa. Using Eq. (3), the fracture toughness $K_{IC}$ of silica is predicted to be in the range of 0.61 – 1.0 MPa√m as the lower and upper limits. Indeed, $K_{IC}$ of amorphous silica thin films in the range of 0.65 – 1.0 MPa√m has been reported [25].

The intrinsic strength $f_t$ of crystalline quartz can be estimated by Eq. (2) with the same theoretical strength $\sigma_{th}$ = 16 – 17 GPa [11,12]. The regular molecular structure of crystalline quartz with $C_{ch}$ = 0.62 nm means the common "intrinsic molecular defect" $a_{th}$ is 0.62 nm, i.e.,

$$f_t = \frac{\sigma_{th}}{\sqrt{1+\frac{a_{th}}{3 \cdot C_{ch}}}} = \frac{16-17}{\sqrt{1+\frac{0.62}{3 \cdot 0.62}}} = 13.8 - 14.7 \ (\text{GPa}) \tag{5}$$

Considering possible variations in the "intrinsic molecular defect" $a_{th}$, the intrinsic tensile strength $f_t$ of crystalline quartz can be adjusted to the range of 13.5 – 15.0 GPa. Then the lower intrinsic strength limit of crystalline quartz is the same as the upper limit of amorphous silica, which is logical for the best possible silica. From the adjusted $f_t$ range, Eq. (3) predicts the fracture toughness $K_{IC}$ of crystalline quartz is in the range of 1.16 – 1.47 MPa√m.

$K_{IC}$ of crystalline quartz has been measured using 7 different indentation formulas under different indentation loads [13], showing $K_{IC}$ varies from 0.5 to 2.5 MPa√m. If the lowest and highest indentation formulas are removed, the reduced $K_{IC}$ range for crystalline quartz is around 1.0 – 1.5 MPa√m. This compares well with the predicted range of 1.16 – 1.47 MPa√m. Predictions from Eqs. (2) and (3) for amorphous silica and crystalline quartz are summarised in Table 1. The predicted fracture toughness $K_{IC}$ is well supported by experiments reported in the literature for both amorphous silica and crystalline quartz.

Table 1. Predictions from Eqs. (2) and (3) based on theoretical strength $\sigma_{th}$ = 16 – 17 GPa [11,12].

| Properties → SiO$_2$ ↓ | Characteristic Molecular Structure $C_{ch}$ (nm) | Intrinsic Strength $f_t$ (GPa) | Fracture Toughness $K_{IC}$ (MPa √m) |
|---|---|---|---|
| **Amorphous Silica** | 0.31 | 10.0 – 13.5 | 0.61 – 1.0 |
| **Crystalline Quartz** | 0.62 | 13.5 - 15.0 | 1.16 – 1.47 |

A silica nano-fibre with the diameter of 1 nm is illustrated in Fig. 2. Because the "molecular crystals" are already 0.31 nm in diameter, the surface of the nano-fibre most likely are not smooth at the atomic scale. Possible unfilled half-circle areas at the nano-fibre surface are illustrated in Fig. 2.

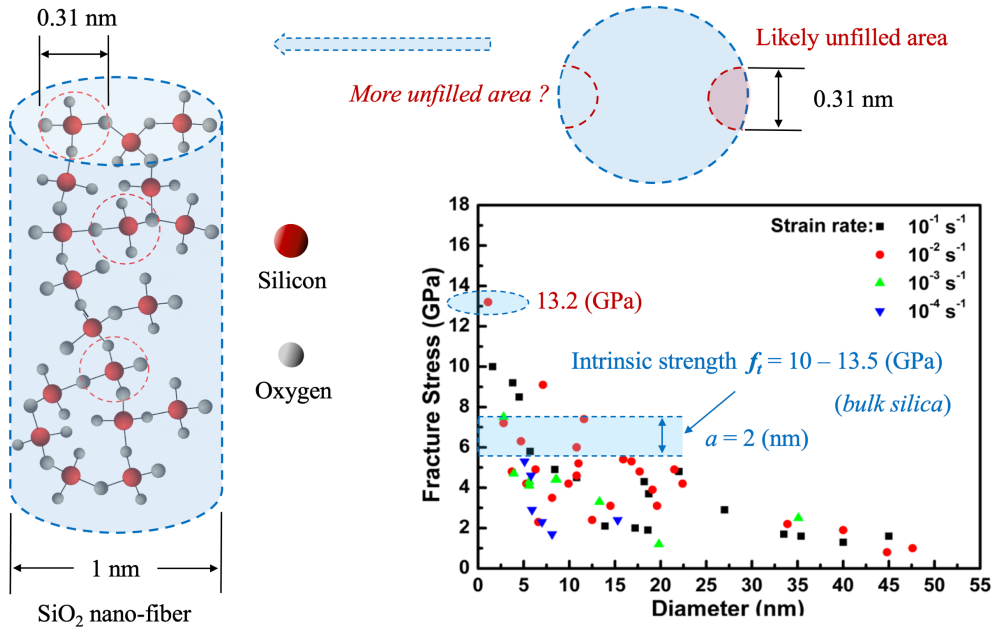

FIG. 2. Schematic 1 nm silica nano-fibre with possible surface defect(s) around 0.31 nm, based on the regular molecular unit or crystal in Fig. 1(f). Measured fracture stress of 13.2 GPa from 1 nm silica nano-fibres, and the predicted fracture stress of 5.6 - 7.6 GPa for bulk silica containing 2 nm defect/crack from Eq. (2).

Due to the limited dimension, the assumed half-circle nano defect in Fig. 2 has to be included in the strength criterion calculation, i.e.,

$$f_t \cdot \left\{\pi \cdot \left(\tfrac{1}{2}\right)^2\right\} \approx \sigma_{th} \cdot \left\{\pi \cdot \left(\tfrac{1}{2}\right)^2 - \tfrac{1}{2} \cdot \pi \cdot \left(\tfrac{0.31}{2}\right)^2\right\} \quad \text{or} \quad f_t = 0.95 \cdot \sigma_{th} \qquad (6)$$

Ignoring the detailed geometry factor for the assumed surface defect in Fig. 2, Eq. (2) can be rewritten as follows, i.e.,

$$\sigma_f = \frac{0.95 \cdot \sigma_{th}}{\sqrt{1 + \frac{0.31}{3 \cdot C_{ch}}}} = \frac{0.95 \cdot (16-17)}{\sqrt{1 + \frac{0.31}{3 \cdot 0.31}}} = 13.2 - 14.0 \text{ (GPa)} \qquad (7)$$

The tensile strength of 13.2 GPa was measured in [11] from the 1 nm silica nano-fibres. If the total unfilled section in the 1 nm silica fibre is around two half-circles, the predicted strength range from Eq. (7) is further reduced by 5% to 12.5 - 13.3 GPa. If bulk silica (e.g. diameter ≥ 10 nm) contains a pre-existing defect/crack of 2 nm, Eq. (2) predicts the fracture stress $\sigma_f$ = 5.6 - 7.6 GPa. Eq. (1b) of LEFM predicts $\sigma_f$ = 6.9 - 11.3 GPa for $a$ = 2 nm and $K_{IC}$ = 0.61 - 1.0 MPa √m.

Eqs. (2) and (3) have also been used for single crystal silicon and diamond [19,20] with the atomic diameter $D_a$ as the characteristic microstructure $C_{ch}$. The estimated $K_{IC}$ for single crystal silicon from a comprehensive review [26] is between 0.75 - 1.30 MPa √m. The theoretical strength $\sigma_{th}$ of single crystal silicon is about 21 GPa for the {111} plane, which is in the middle range of theoretical strengths of different crystal planes [27]. The atomic diameter of silicon $D_a$ = 0.235 nm [28,29]. Eq. (3) predicts the mid-range $K_{IC}$ value of single crystal silicon is about 1.12 MPa √m. Furthermore, the model has been verified by nano-ceramics [30], polycrystalline ceramics, rocks and concrete [20]. The range of the characteristic microstructure $C_{ch}$ can vary from 0.1 nm to close to 100 mm, or with the size ratio close to $10^9$.

Following the above discussions, Eq. (3) can also be applied to nanocrystalline high toughness SiO₂ stishovite [31,32]. For instance, it was reported in [32] that the grain size $C_{ch}$ = 128 nm and the fracture strength $\sigma_f$ = 6.3 GPa for 0.5h samples. It is relatively easy to suppress the influence of nano-defects with careful sample preparations because of this large nano-grain size of 128 nm. In this case, it can be assumed that the fracture strength $\sigma_f$ is about the same as the intrinsic tensile strength $f_t$. Then Eq. (3) predicts that this high toughness SiO₂ stishovite has $K_{IC}$ = 7.8 MPa√m. The crack growth resistance $K_R$-curve is between 3.5 MPa√m to around 9-10 MPa√m for crack growth < 5 $\mu m$ [32]. The predicted fracture toughness $K_{IC}$ of 7.8 MPa√m is well within the $K_R$-curve range.

It should be emphasized that the possibility for accurate predictions of the fracture toughness $K_{IC}$ based on the atomic, molecular and nano-grain structures as presented in this study is significant, because previously it is believed that the bulk or macroscopic fracture toughness $K_{IC}$ can only be measured experimentally using samples with man-made cracks. Eq. (3) provides an alternative way for determination of $K_{IC}$ based on the theoretical (or intrinsic) strength and characteristic microstructure $C_{ch}$.

The important relation between the intrinsic tensile strength $f_t$ and characteristic microstructure $C_{ch}$ can be defined by Eq. (3). Rearranging the formula, it can be obtained that:

$$f_t = 0 + \frac{K_{IC}}{2\sqrt{3}} \cdot \frac{1}{\sqrt{C_{ch}}} \tag{8}$$

The Hall-Petch relation [33-35] linking the yield strength $\sigma_Y$ of a metal to its average grain size $d_G$ has the following form.

$$\sigma_Y = \sigma_0 + k \cdot \frac{1}{\sqrt{d_G}} \qquad (20\ nm\ <\ d_G\ <\ 100\ \mu m) \tag{9}$$

The comparison of Eqs. (8) and (9) shows that Eq. (3) or (8) is the "Hall-Petch relation" for brittle solids [19]. The difference is Eq. (9) is not predictive as the two fitting parameters, $\sigma_0$ and $k$, need to be determined experimentally for a give metal. Also, Eq. (3) or (8) is simpler than the classic Hall-Petch relation as $\sigma_0 = 0$ and $k$ is well defined by the fracture toughness $K_{IC}$. As a result, Eq. (3) or (8) is predictive. Both Eqs. (8) and (9) are illustrated schematically in Fig. 3 for a wide range of the characteristic microstructure $C_{ch}$ or grain size $d_G$.

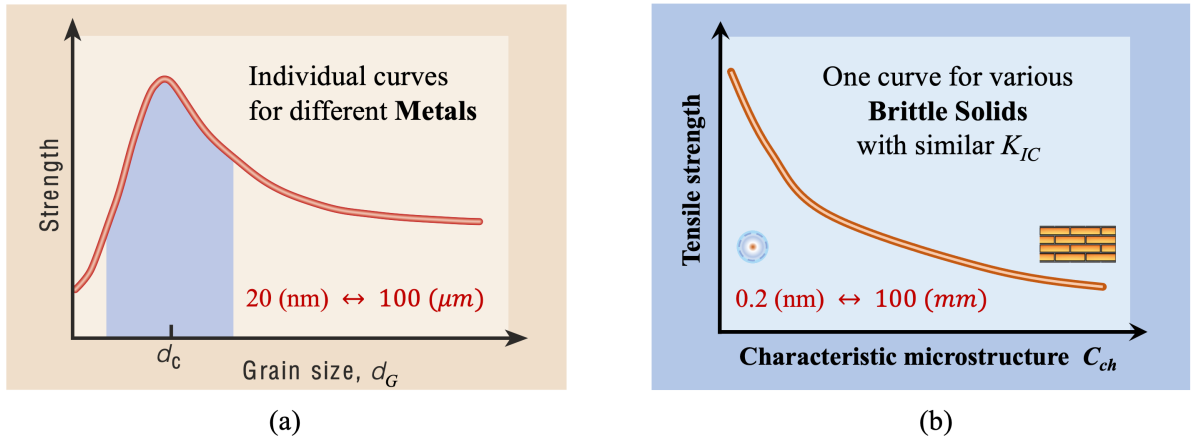

FIG. 3. (a) Hall-Petch relation for metals [36], one fitted curve for one metal only. The critical grain size $d_c$ for the transition is around 20 (nm) for the anti-"Hall-Petch" effects. (b) Monotonic "Hall-Petch" relation for brittle solids with similar fracture toughness $K_{IC}$ [19].

Science 2021 compiled 125 questions: Exploration and Discovery [37]. One of the Physics questions is "Can we accurately simulate the macro- and microworld?". It has been shown in this study that the macroscopic fracture toughness and intrinsic strength of amorphous silica and crystalline quartz can be accurately predicted by their molecular structures or characteristic molecular crystals with $C_{ch}$ = 0.31 and 0.62 nm, together with the theoretical strength $\sigma_{th}$ (linked to the atomic bonds). Eqs. (2) and (3) are simple predictive formulas. The comparison of Eqs. (8) and (9) is significant as Eq. (3) in fact is the "Hall-Petch relation" for brittle solids [19] with the characteristic microstructure $C_{ch}$ varying from atomic and molecular scale, around 0.31 to 0.62 nm for silica and quartz, to nano-scaled $SiO_2$ stishovite around 100 nm, and then all the way to micro-and macro-scaled microstructures of various solids [20]. That is the quasi-brittle fracture behaviour requiring non-LEFM modelling can be observed in all brittle solids at the scale of their microstructures. Therefore, it is correct to say "Glass Breaks like Metal, but at the Nanometer Scale" [22].

---------------------------